\documentclass[preprint,showpacs,aps]{revtex4}

\usepackage{graphicx}% Include figure files
\usepackage{dcolumn}% Align table columns on decimal point
\usepackage{bm}% bold math

\usepackage{graphicx}% Include figure

\begin{document}

\title{Transport and dynamical properties of inertial ratchets}% Force line breaks with \\

\author{M.F.Carusela}

\altaffiliation[ ]{Consejo Nacional de Invesigaciones Cientificas y Tecnicas,CONICET.}
\email{flor@ungs.edu.ar} 
 \affiliation{Instituto de Ciencias, Universidad de General Sarmiento, 
J.M.Gutierrez 1150, (1613) Los Polvorines, Buenos Aires, Argentina.}%Lines break automatically or can be forced with \\
\author{A.J.Fendrik}% 
\altaffiliation[ ]{Consejo Nacional de Invesigaciones Cientificas y Tecnicas,CONICET.}%Lines break automatically or can be forced with \\
\email{fendrik@df.uba.ar} 
\affiliation{Departamento de F\'{\i}sica J.J.Giambiagi, Facultad de Ciencias Exactas y Naturales. Universidad de Buenos Aires.(1428) Buenos Aires. Argentina.
}%

\author{L.Romanelli}
\altaffiliation[ ]{Consejo Nacional de Invesigaciones Cientificas y Tecnicas, CONICET.}%Lines break automatically or can be forced with \\
\email{lili@ungs.edu.ar}
\affiliation{Instituto de Ciencias, Universidad de General Sarmiento, 
J.M.Gutierrez 1150, (1613) Los Polvorines, Buenos Aires, Argentina.
% with \\
}%

\date{\today}% It is always \today, today,
             %  but any date may be explicitly specified

\begin{abstract}
In this paper we discuss the dynamics and transport properties of a massive particle, in a time dependent periodic potential of the ratchet type, with a dissipative environment. The directional currents and characteristics of the motion are studied as the specific frictional
coefficient varies, finding that the stationary regime is strongly dependent on this parameter. The maximal Lyapunov exponent and the current show large fluctuations and inversions, therefore for some range of the control parameter, this inertial ratchet could originate a mass separation device. Also an exploration of the effect of a random force on the system is performed.    

%PACS numbers may be entered using the \verb+\pacs{#1}+ command.
\end{abstract}

% PACS, the Physics and Astronomy
                             % Classification Scheme.
%\keywords{Suggested keywords}%Use showkeys class option if keyword
                              %display desired
\maketitle

\section{Introduction}
\label{sec:intro}

Transport phenomena play a crucial role in many problems from physics, biology and social science. In particular, there have been an increasing interest in transport properties of the Brownian ratchets (BR), devices out of thermal equilibrium in which a nonzero net drift velocity may be obtained from fluctuations interacting with broken symmetry structures\cite{denisov,flach}. These devices that where first proposed by Smoluchowsky \cite{smolu} 
and later discussed by Feynmann \cite{feyn} have deserved a great deal of attention in the 
literature (for a review see Ref.\cite{reimann}). There 
is a wide diversity of areas in which BR's are applied, for instance to the study of molecular 
motors \cite{astumian}, the description of ion channels and molecular transport within 
cells \cite{bier}, the treatment of Parrondo's paradoxical games \cite{parrondo,harmer}, optical ratchets and directed motion of laser cooled atoms \cite{mennerat} and coupled Josephson junctions \cite{zapata}. In particular, these nonequilibrium models become especially interesting due to their technological applications on nanoscales and microscales \cite{ajdari,kaplan} such as microscopic particle separation \cite{ertas,duke,lindner}.

In these models a net transport can be induced with fluctuations associated with an additive force and noise, but also without noise, in overdamped \cite{popes} or underdamped systems \cite{mateos,hangui}. In particular, inertial ratchets may exhibit a very complex and rich dynamics, both regular and chaotic motion. Moreover, they can display multiple reversal currents where their directions depend on the inertia term and the amount of friction.

In this paper, the problem of transport in periodic asymmetric potential of the ratchet type is addresed. We study a model 
with an underdamped ratchet under the influence of spatio-temporal external perturbations without and with noise. 
Other authors, have studied the reversal current as a function of the intensity of an external driving \cite{mateos,hangui,larrondo}.
One of the facts of this paper is to reveal the current reversal as a function of the specific frictional coefficient that provides a mass separator device.

The present work is organized as follows: in Section \ref{sec:dos} we introduce the model
under study; in Section \ref{sec:tres} the dynamical study of the system in particular in the transport regime is presented. Section \ref{sec:cuatro} is devoted to the study of the role of noise to modify the characteristis of the dynamics and to induce the transport current . Finally in Section \ref{sec:cinco} we summarize our results. Some of  technical details are given in an Appendix.

\section{Anular inertial ratchet}
\label{sec:dos}

\begin{figure}
\includegraphics{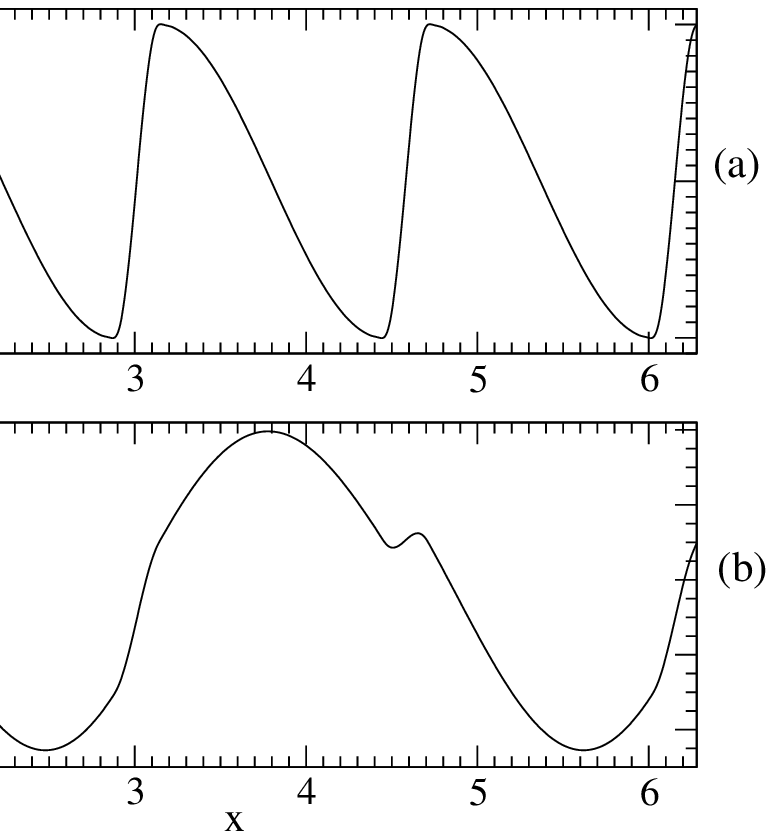}
\caption{\label{fig:figuextra}} 
(a) Static potential $V_{\alpha}(x)$ with $\alpha=5$ as a function of the coordinate $x$.
(b) Total potential $V_{\alpha}(x)+V^{dr}(x,t)$ with $\varepsilon<\varepsilon_{c}$ for $t=\pi/2\omega$ as a function of the coordinate $x$.
\end{figure}

We consider non interactive massive ($m$) particles placed in a one-dimensional ring of radius $R=1$, inmmersed in a dissipative environment, under a periodic potential and driven by an external force. The time evolution is given by the equation:
\begin{equation}
\label{dinamica}
m \ddot{x}=-\mu \dot{x}-\frac{\partial V_{\alpha}(x)}{\partial x}+F^{dr}(x,t).
\end{equation}
In Eq.[\ref{dinamica}], $0 \leq x(t) \leq 2\pi$ represents the coordinate of the particle, $\mu$ the damping coefficient and $V(x)_{\alpha}$ is a one dimensional, periodic potential given by
%Arreglar el potencial
\begin{equation}
\label{eq:exacto}
V_{\alpha}(x)=\left\{
\begin{array}{ll}
V_{o} \cos\left(\frac{\pi (\alpha+1) x}{a \alpha} \right), \;
\mbox{for} \; 0\leq x  \leq \frac{a \alpha}{(\alpha+1)}\\ 	                    
-V_{o}\cos\left(\frac{\pi (\alpha+1) x}{a}-\alpha \pi \right),\; \mbox{for}\; \frac{a \alpha}{(\alpha+1)}\leq x\leq a , 
\end{array}\right.
\label{pot}
\end{equation}
fulfilling the periodicity condition $V_{\alpha}(x+a)= V_{\alpha}(x)$ where $a=2\pi/N$ 
being $N$ the number of wells (or sites) along the ring. The parameter $\alpha$ ($\alpha>0$) controls the left-right asymmetry of the potential. 
For $\alpha >1 (<1)$ the minimum in each well of the ratchet is displaced towards the right (left) while $\alpha=1$ corresponds to left-right symmetric potential. Figure \ref{fig:figuextra}(a) displays this static potential for $N=4$ and $\alpha=5$.

Particles are driven by the external periodic driving force $F^{dr}(x,t)$. We take this to be the gradient of a time dependent potential with a spatial periodicity that is twice the one of $V_{\alpha}$ in order that 
consecutive wells alternate in time as absolute minima. This condition restricts $N$ to be even. We therefore consider
\begin{equation}
\label{fuerza}
F^{dr}(x,t)=-\varepsilon \frac {\partial V^{dr}(x,t)}{\partial x}=-\varepsilon \sin (\omega t)\frac {d \sin ( N x/2)}{dx} ,
\end{equation}
that represents a longitudinal stationary wave along the ring. 
Time is measured in units of the period $\tau = 2\pi/\omega$ of the external driving and $\varepsilon$ is the coupling strength. There is a critical value $\varepsilon=\varepsilon_{c}$ such that if $\varepsilon<\varepsilon_{c}$ the whole potential ($ V_{\alpha}(x)+V^{dr}(x,t)$) always has a minimum per site. For $\varepsilon>\varepsilon_{c}$ the potential may loose minima corresponding to alternate sites as the time varies. Figure \ref{fig:figuextra}(b) displays the whole potential at $\tau/4$ for $N=4$ and $\alpha=5$ and $\varepsilon<\varepsilon_{c}$.

According the values of the parameters and the initial conditions the motion would be 
bounded (libration near the sites) or unbounded (rotations around the ring).

To obtain a directional current in the unbounded dynamics, it is useful to study the symmetries of the equation of motion Eq.[\ref{dinamica}]. For hamiltonian systems (i.e. $\mu=0$ ) with $\alpha=1$, there are two symmetries that prevent directional transport:
\begin{eqnarray}
S_{1}: x \rightarrow x+(2k+1)\frac{2\pi}{N}, \; t \rightarrow -t, \; v \rightarrow -v ;\\
S_{2}: x \rightarrow (2k+1)\frac{2\pi}{N}-x, \; t \rightarrow t, \; v \rightarrow -v.
\end{eqnarray}
$S_{1}$ symmetry leaves invariant Eq.[\ref{dinamica}] while $S_{2}$ changes the sign the whole expression. In accord to both symmetries, for each solution with velocity $v$ will be another solution with value $-v$. If $\alpha \not= 1$ $S_{2}$, symmetry is removed but $S_{1}$ still holds. When $\mu>0$ the only symmetry removed is $S_{1}$. Therefore, for obtaining directional transport the values of $\alpha$ and $\mu$ have to be $\alpha \not=1$ and $\mu>0$.
 Taking this into account, we claim that the current is induced by damping. In the overdamped limit, such that the inertial term can be dropped out, the transport phenomena in the ratchet described by Eq.[\ref{dinamica}] (setting $m=0$) has been well understood
\cite{nos1}. To determine inertial effects on the directional current, we will study the characteristics of the dynamics as a function of the mass.

For practical reasons, we have numerically solved Eq.[\ref{dinamica}] using the Fourier expansion of the potencial $V_{\alpha}(x)$ Eq.[\ref{pot}] up to 20-th order rather than the original one (see appendix).

\section{Transport and dynamics.}
\label{sec:tres}

In this section we will discuss the characteristics of the directional transport and the dynamics of the ratchet as a function of the specific frictional coefficient $\mu/m$. Let us consider the dimensionless
version of Eq.[\ref{dinamica}] as: 
\begin{equation}
\label{adim}
\ddot{x}=\beta \left[- \dot{x}+\gamma \left( F_{\alpha}(Nx)-\frac{\varepsilon^\prime}{2}\cos{(Nx/2)} \sin{(2\pi t)}\right)\right] ,
\end{equation}
where $\beta=\frac{2\pi\mu}{m \omega}$ is the specific frictional parameter, $\gamma=\frac{2\pi V_{o} N}{\mu \omega}$ is the ratio between the external force and the dissipation, 
$F(Nx)=-\frac{1}{N V_{o}}\frac{\partial V_{\alpha}(N x)}{\partial x}$ and $\varepsilon^\prime=\varepsilon/V_{o}$ (in the following we drop the prime).

All the calculations that we report where made for $N=4$, $V_{o}=10$, $\omega =6$, 
$\mu=1$ (that is $\gamma=40\pi/3$) and $\varepsilon =6.5$ greater than $\varepsilon_{c}=6.26007$. 
We fix all the parameters except $m$, therefore we study the system as long as $\beta$ is changing.

We study the dynamical behavior of the system in the stationary regime starting from an ensamble of one hundred random initial conditions ($N_{in}=100$) for each $\beta$ . We have calculated the
the mean velocity $\bar{v}$, measure in site per period of the driving force, for each initial condition is:
\begin{equation}
\bar{v}= \left(\frac{N}{2\pi}\right) \lim_{t\to\infty} \frac{(x(t)+2\pi k)}{t} ,
\end{equation}
where $k$ is the winding number (that is the number of rotation that the particle makes
around the ring).
Then, the average ensamble velocity is given by:
\begin{equation}
\langle \bar{v} \rangle = \frac{1}{N_{in}} \sum_{j=1}^{N_{in}} \bar{v}_{j} ,
\end{equation}
and its fluctuations:
\begin{equation}
\sigma_{\langle \bar{v} \rangle} = \sqrt{\frac{1}{N_{in}} \sum_{j=1}^{N_{in}} (\bar{v}_{j}-\langle \bar{v} \rangle)^{2}} .
\end{equation}

In the same way, the average maximum Lyapunov exponent is:
\begin{equation}
\langle L_{max} \rangle = \frac{1}{N_{in}} \sum_{j=1}^{N_{in}} L_{max}^{j} ,
\end{equation}
where $L_{max}^{j}$ corresponds to the maximum Lyapunov exponent for the $j-th$ initial condition (calculated by algorithm given in Ref.\cite{lyapurui}) and the fluctuation:
\begin{equation}
\sigma_{\langle L_{max} \rangle} = \sqrt{\frac{1}{N_{in}} \sum_{j=1}^{N_{in}} (L_{max}^{j}-\langle L_{max} \rangle)^2} .
\end{equation}

\begin{figure}
\includegraphics{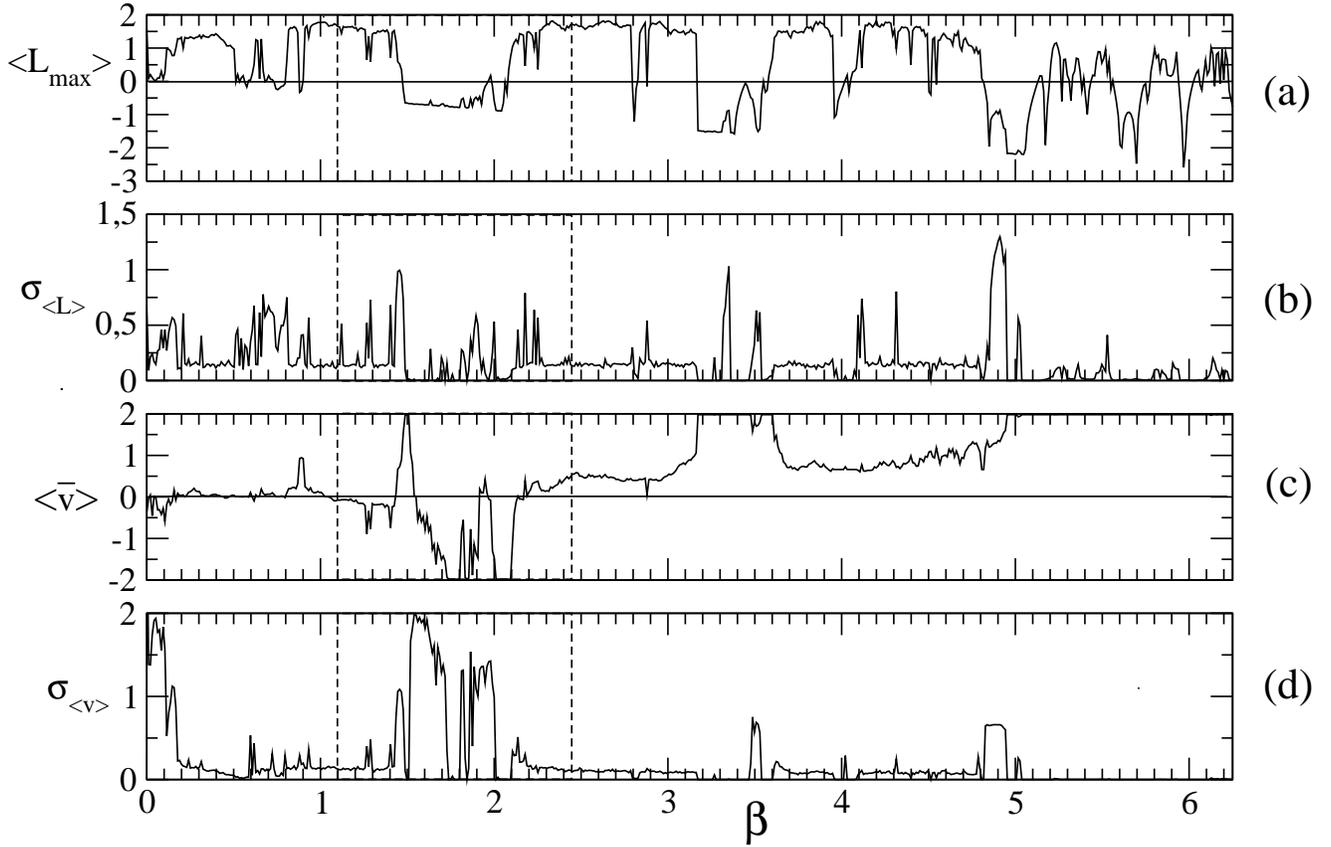}
\caption{\label{fig:figura1} 
Average of the maximum Lyapunov exponent $\langle L_{max}\rangle$ (a), dispersion of the maximum Lyapunov exponent $\sigma_{L}$ (b), average velocity $\langle \bar{v} \rangle $ (c) and dispersion of average velocity  $\sigma_{\langle v \rangle}$ (d) as a function of the dissipation $\beta$. The velocity is measured in terms of sites per period of the driven force. For sake of clarity in the four graphs the points are linked by lines. The vertical dashed lines limit the region shown in Fig.\ref{fig:figura2}.}
\end{figure}

\begin{figure}
\includegraphics{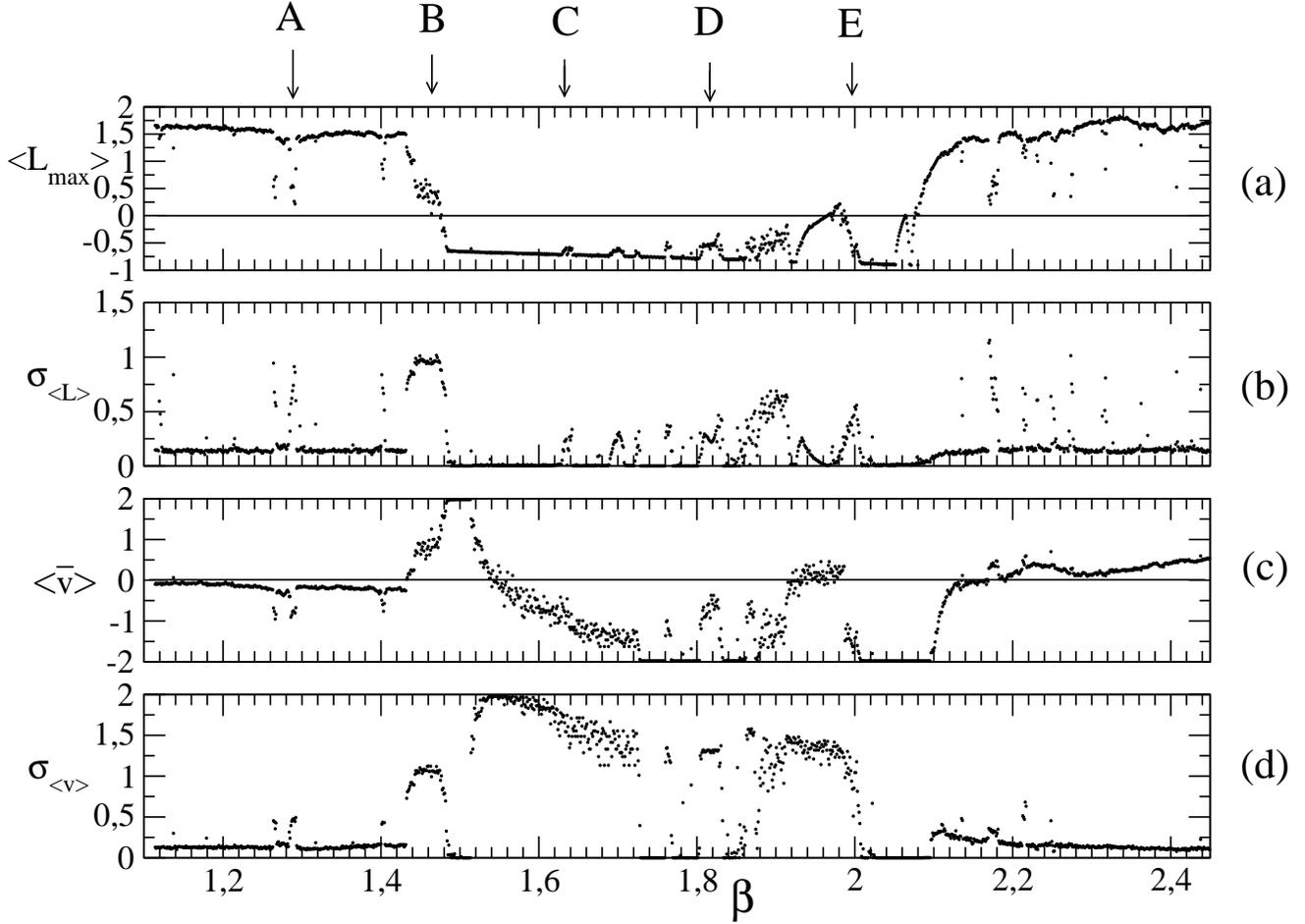}
\caption{\label{fig:figura2}
Blow up of Fig. \ref{fig:figura1}.
 }
\end{figure}

In Figure \ref{fig:figura1} the magnitudes defined above are displayed. By inspecting this
figure (a,b,c, and d), we found diverse and complementary information that can be extract from them. From Figure \ref{fig:figura1}(a), we observe the average of the maximum Lyapunov exponent belonging to the mentioned stationary orbits. These exponents have positive and negative values. The negative values, as usual, would be related with regular behavior meanwhile the positive with irregular one, this assert holds when the dispersion shown in Fig.\ref{fig:figura1}(b) remains small.
However from the same figure, we found coexistence of different attractors. We discuss some of them, indicated by the arrows in Fig.\ref{fig:figura2}.
The average ensamble velocity is shown in Fig.\ref{fig:figura1}(c) while in Fig.\ref{fig:figura1}(d)
its dispersion. It can be observe for $\beta>5$ an average current $\langle \bar{v} \rangle_{max}=2$ is established with zero dispersion. We remark that in such limit, the system is overdamped and 
the inertial term is negligeable for studying the current (see ref.\cite{nos1}) althougth its effect is present in the Lyapunov exponent as can be seen from its fluctuations above $\beta=5$ (Fig.\ref{fig:figura2}) related to the intrawell dynamics. 
For lower values of $\beta$, a complex behavior is observed: the values of the current fluctuates from $-2$ to $2$ in some regions while it vanishes in others. The fluctuations displayed in the figures do not correspond to a statistical effect but to the complex structure of the dynamic (i.e. appearance or desappearance of diferent kind of orbits), what is therefore worth to a more detailed analysis. We varied $\beta$ with a smaller step in order to get a zoom in a particular range as can be seen in Fig.\ref{fig:figura2}.   
The arrows show particular values where there are coexistence of attractors. The label A
correspond to a chaotic orbit (with almost no transport) together with a regular one whose
current is $\bar{v} =-4/3$. B is a region where arise a regular orbit of current $\bar{v}=2$ meanwhile
the chaotic orbit persists near to $\beta=1.49$. In C three regular orbits with current 
$\bar{v}=2$, $\bar{v}=-2$ and $\bar{v}=0$ are present. The label D corresponds to coexistence of two regular orbits of currents $\bar{v}=-2$ and $\bar{v}=2/3$ respectively.  
 
\begin{figure}
\includegraphics{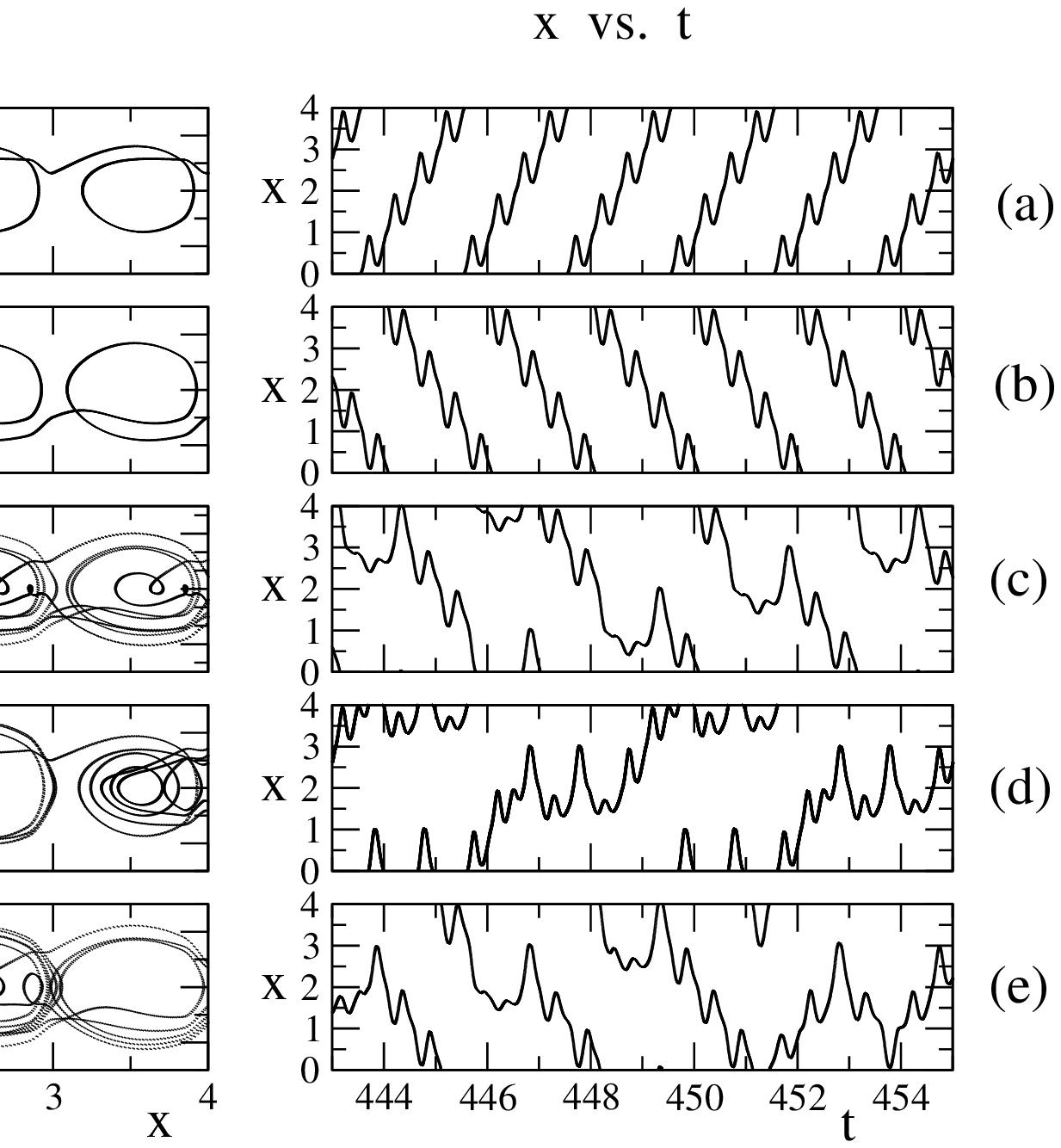}
\caption{\label{fig:figura3}
Projection of the trajectory in the phase space and the position of the particle versus time for differents values of $\beta$. (a) Regular orbit with mean velocity $\bar{v}=2$. (b) Same as above with $\bar{v}=-2$. (c) $\bar{v}=-4/3$. (d) $\bar{v}=2/3$ and (e) chaotic behavior with $\bar{v}=0$. 
 }
\end{figure}
Figure \ref{fig:figura3} displays the position of the particle as a function of time and the projection of the trajectory onto the phase space. The cases shown are related to those cases discussed in the text. 
\begin{figure}
\includegraphics{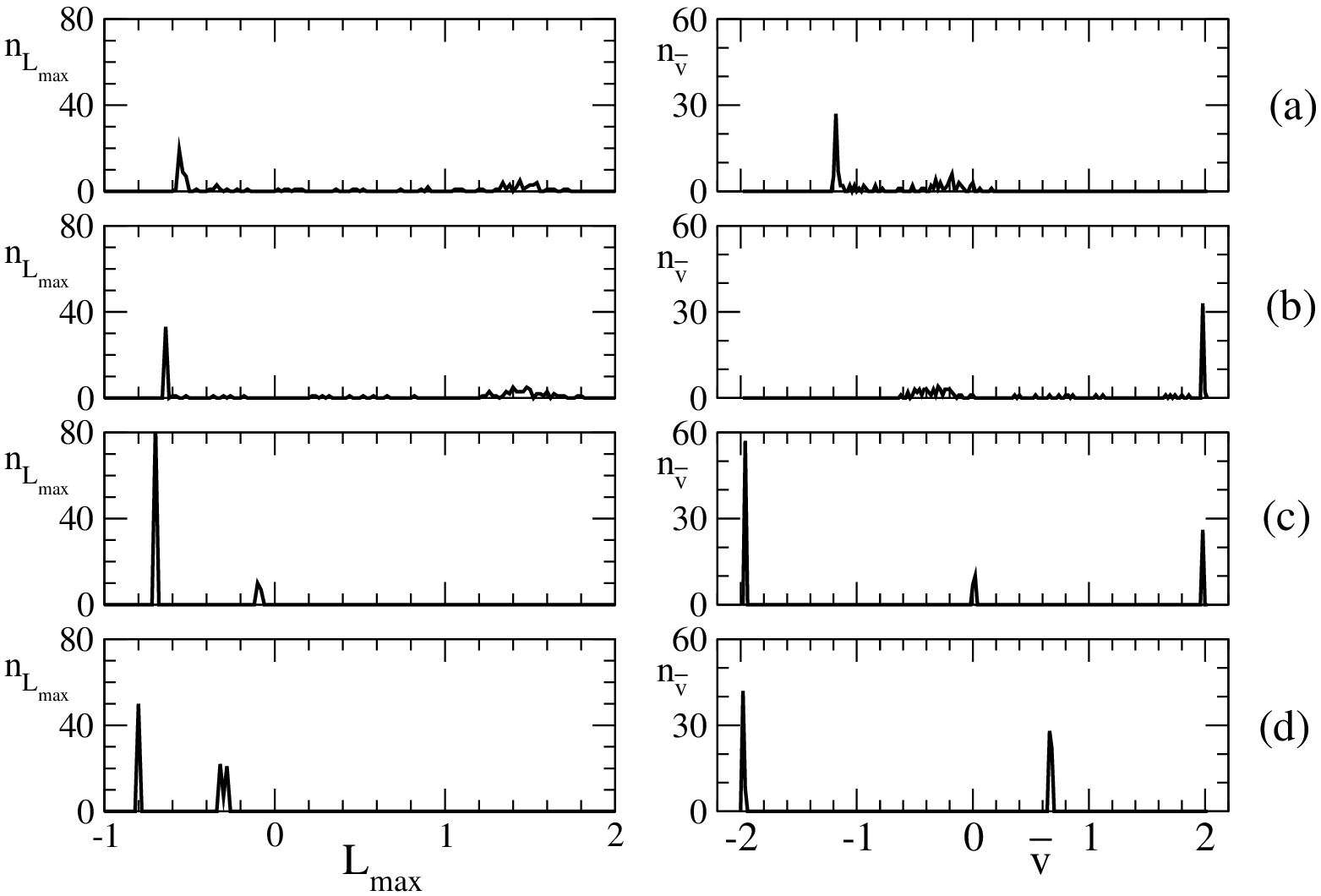}
\caption{\label{fig:figura4}
Histogram of the maximum Lyapunov exponent (left panel) and histogram of the mean velocity
$\bar{v}$ (right panel) for different values of $\beta$. All the calculations where obtained with an ensamble of 100 random initial conditions. Each row (from top to bottom) corresponds to the regions indicated by arrows A, B, C and D in Fig. \ref{fig:figura2}.
 }
\end{figure}

To stress the coexistence of attractors for a given damping value, as discussed above, the histograms of the maximun Lyapunov exponent ($n_{L_{max}}$) and the mean velocity ($n_{\bar{v}}$) are displayed in Fig.\ref{fig:figura4}. The highest value of $n_{L_{max}}$ in the row (c) is the sum of the values of $n_{\bar{v}}$ for $\bar{v}=2$ and $\bar{v}=-2$ because both orbits have the same maximal Lyapunov for this $\beta$.

It is interesting to note that the current $\langle \bar{v} \rangle$ in Fig.\ref{fig:figura1}(c) shows abrupt inversions as a $\beta$ changes in a similar way to other ratchets when the parameters varies. As the variation of $\beta$ depends only on the mass $m$, similar particles
but different masses will have different velocities. Therefore, this behavior originates a
mass separation device.

\section{The effect of noise}
\label{sec:cuatro}

We now turn to study of transport current assisted by noise in presence of an external random force $\xi(t)$ fulfilling $\langle \xi(t) \xi(t') \rangle =\sigma^{2} \delta(t-t')$. We solve the following stochastic differential equation:
\begin{equation}
\label{adimrui}
\ddot{x}=\beta \left[- \dot{x}+\gamma \left( F_{\alpha}(Nx)-\frac{\varepsilon^\prime}{2}\cos{(Nx/2)} \sin{(2\pi t)}\right)+\xi(t) \right] .
\end{equation}
 
In the noiseless overdamped limit ($m \rightarrow 0$), the motion is bounded ($\bar{v}=0$) or unbounded 
($\bar{v} \neq 0$) according to $\varepsilon < \varepsilon_{c}$ or 
$\varepsilon > \varepsilon_{c}$. In both cases the stationary motion is a single regular orbit robust under slight variation of the coupling. As it was shown in ref. \cite{nos1} the addition of an appropriate noise to the bounded regular orbit can induce a transition to the unbounded motion ($\bar{v}\neq 0$) rather than its destruction. In other words, the effect of noise on the bounded motion can mimic the increase of coupling $\varepsilon $ from a lower to a higher value than the critical, therefore we will have a brownian ratchet whose maximal current is near to those of the unbounded orbit.

This scope would be quite different when the inertial term is non negligeable since the stationary unbounded motion is now more complex due to the fact that small variation in the coupling may introduce or destroy orbits \cite{larrondo}. On the other hand in the inertial system the bounded and the unbounded motion are separated by a fuzzy region of $\varepsilon$ 
where there is coexistence of both types of motion. 

\begin{figure}
\includegraphics{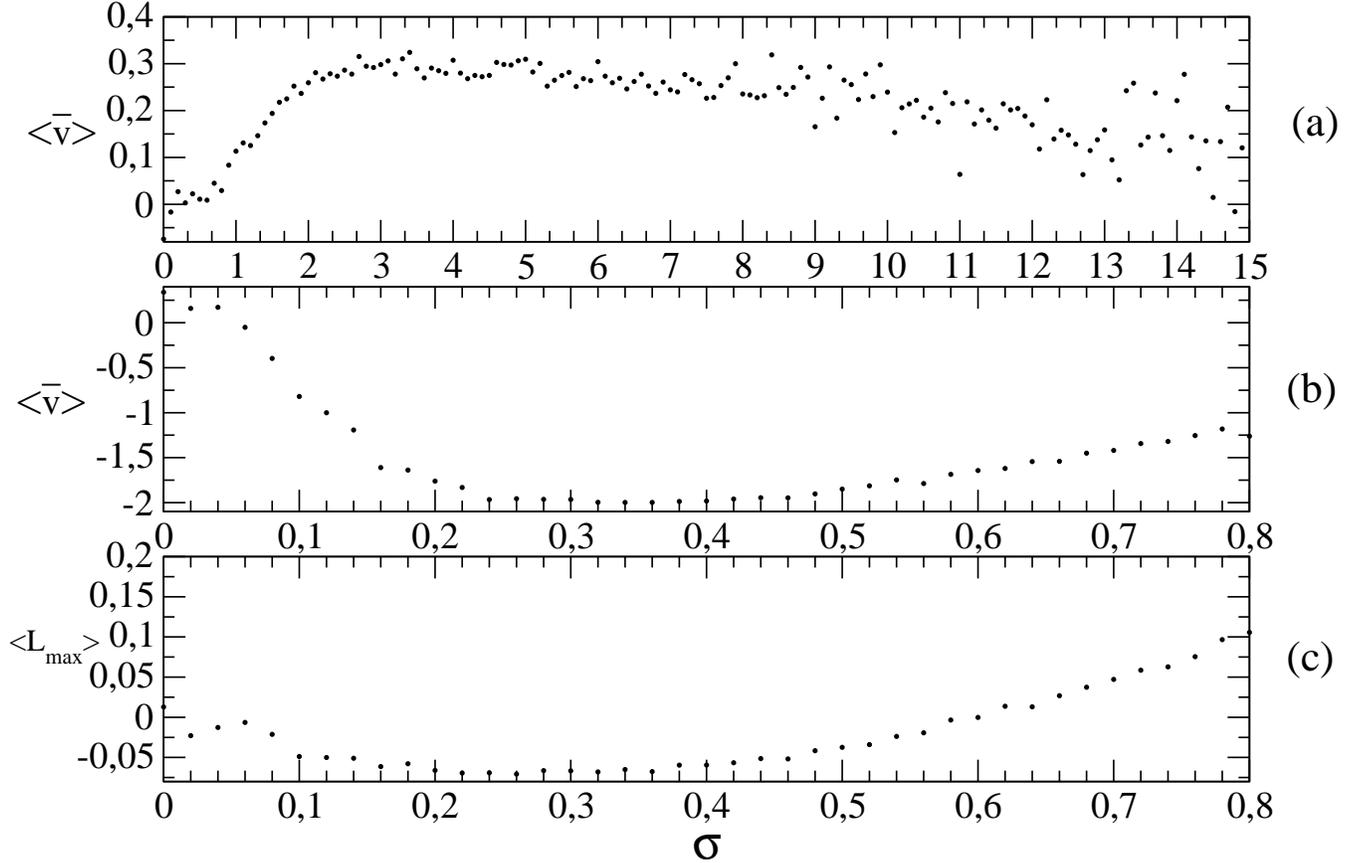}
\caption{\label{fig:figura5}
a) Average velocity  $\langle \bar{v} \rangle$ vs. $\sigma$. The chosen parameters are $\varepsilon=4.1$,
$\omega=6$, $\beta=1.5708$. The calculations where obtained with an ensamble of 200 random initial conditions for each value of $\sigma$.
b) Average velocity  $\langle \bar{v} \rangle$ vs. $\sigma$. The chosen parameters are $\varepsilon=6.5$, $\omega=6$, $\beta=1.987$.
c) Average of the maximal Lyapunov $\langle L_{max} \rangle$ as a function of $\sigma$.
The parameters are the same as in b).}
\end{figure}

Unlike the overdamped system the mere addition of noise to the bounded motion does not lead to obtain transport although it depends onto the robustness and the mean velocity of the unbounded orbits in the chosen parameter region ($\beta, \gamma, \varepsilon$) therefore it is not easy to predict when there will be directional currents. As an example, in Fig.\ref{fig:figura5}(a), we show the current obtained as a function of noise added to a bounded system ($\varepsilon=4.1, \beta=1.5708$). In this case, the maximum value of the velocity corresponds to an optimal noise (near $\sigma=3.5$) but its value is far from
the maximum value which is reached in the overdamped system.

\begin{figure}
\includegraphics{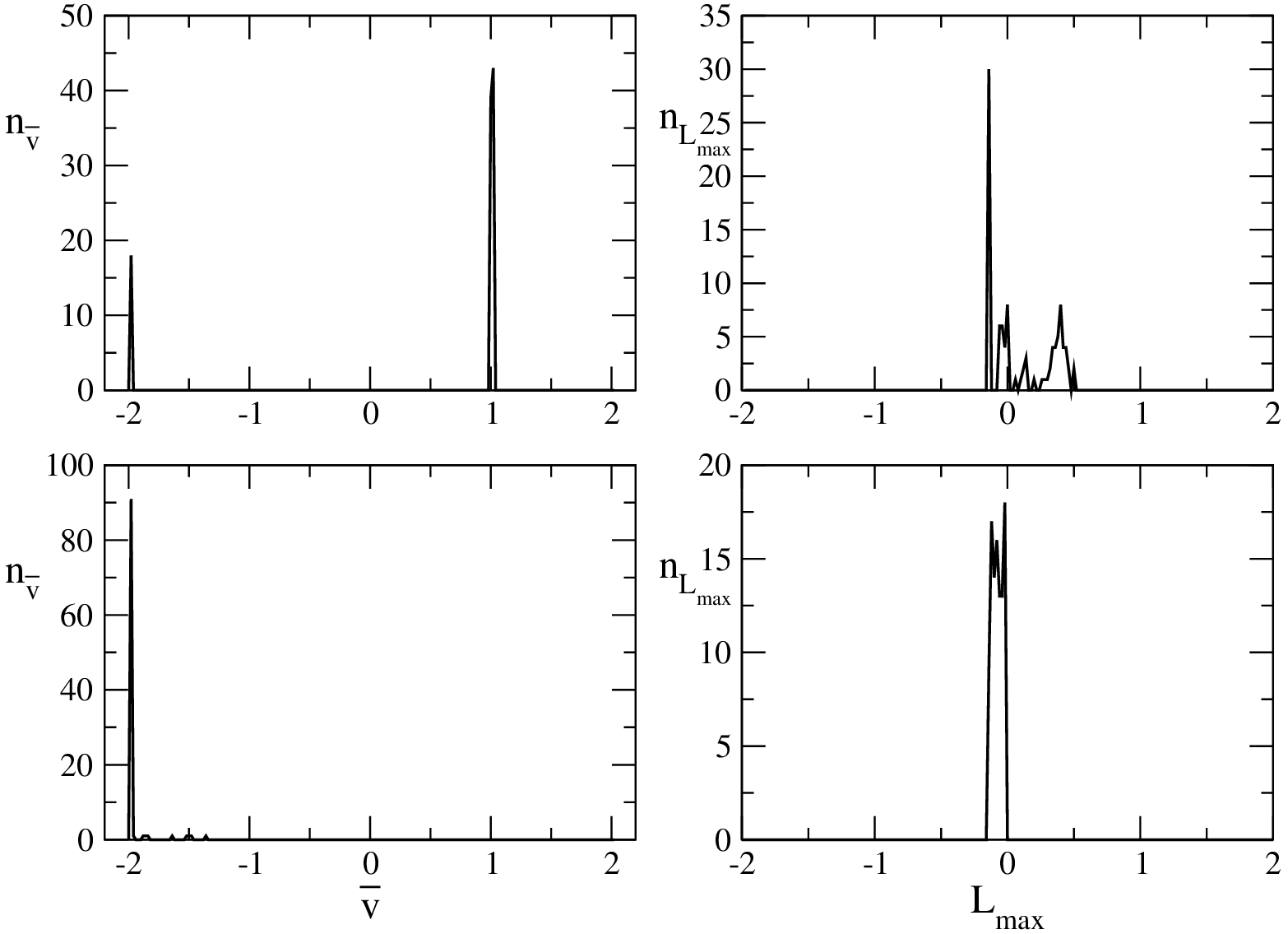}
\caption{\label{fig:figura6}
The upper row displays the histograms of the mean velocity $\bar{v}$ and maximal Lyapunov exponent $L_{max}$ for deterministic systems for $\beta=1.987$, $\varepsilon=6.5$.
The lower row shows the same histograms when the random force $\sigma=0.2$ is added.
 }
\end{figure}

Now, we add the random force to a system where should have transport even in absence of noise (as that shown in \ref{fig:figura2}). The effect of noise will strongly depend upon the parameter $\beta$ because, as we have already seen, it determines the properties of the transport in the stationary regime (current of the single orbits, coexistence of orbits and their stability etc.). Let us consider the deterministic system whose $\beta=1.987$ is indicated by the arrow E in the Fig.\ref{fig:figura2}. There are coexistence of two orbits. One of them, has current $\bar{v} =-2$ and slightly negative maximal Lyapunov exponent and the other one has current $\bar{v} =-1$ and slightly positive maximal Lyapunov exponent. In the first row of Fig. \ref{fig:figura6} the histograms of both are displayed. The mixing of two orbits ocurs in such a way that the total average current and the average maximal Lyapunov are $\langle \bar{v} \rangle=0.34$, $\langle L_{max}\rangle=0.013$ respectively. 

When the random force is added,   
for an optimal noise ($\sigma=0.3$), the stationary orbits have $\bar{v} =-2$ and slightly negative maximal Lyapunov exponents. This can be observed in Fig.\ref{fig:figura5}(b) and (c) where $\langle \bar{v} \rangle$ and $\langle L_{max}\rangle$ as a function of noise
$\sigma$ are displayed. For sake of clarity, in the second row of Fig.\ref{fig:figura6} the histograms of $\bar{v} =-2$ and $\langle L_{max}\rangle$ (see Ref.\cite{lyapurui}) are displayed when the noise is near of the optimal one.
 
\section{Summary and conclusions}
\label{sec:cinco}

In this work we have studied the dynamics in the stationary regime of an inertial ratchet as a function of the specific frictional coefficient ($\beta$). This dynamics is strongly dependent on such coefficient since in some regions slight variation of this parameter can destroy stationary orbits and/or it can create new ones. There are regions where two or three orbits of different characteristics coexist. This can be verify from the fluctuations of the maximal average Lyapunov exponent as long as the respectively histograms. This fact is reflected in the transport properties of the system, since the mean velocity of the orbits involved, should be quite different not only in their magnitude although in their directions. The large fluctuations and inversions showed in the average velocity when $\beta$ varies are the manifestation of such phenomenum. This property could be useful to build a mass separation device.

We have also explored the effect of noise in the system when a random force is added.

In the overdamped limit, when the deterministic dynamics is robust enough against noise
it is possible to obtain a directional current assisted by noise by adding a random force to the system when its motion is bounded. In other words, the presence of noise in moderate amounts induces the transitions bounded-unbounded (libration-rotation) before the destruction of the deterministic dynamics structure. If the inertial term becomes relevant, 
our results show that it is possible to obtain this effect but with a maximal current lesser than the ideal. This behavior may be due to two different causes: 1) The presence of the inertial term makes the deterministic dynamics less robust against the noise 2) The unbounded orbits where the system access by adding the noise have smaller values of $\bar{v}$ than the ideal $\bar{v}$ for the choosen parameters. This could be decided with a more exhaustive analysis which is beyond the scope of the present work.

Next, we explore the inclusion of a random force in the deterministic system where the dynamics is already unbounded. We have consider a region where there are coexistence of two stationary orbits with different characteristics (in particular their $\bar{v}$ have opposite sign such that the average $\langle \bar{v} \rangle$ is slightly positive). We observe, when noise is added, one orbit is not robust and it dessapears while the other one (with $\bar{v}=-2$) persists. Then, the current of the system, not only has an inversion but increases its absolute value. We wish to emphasize that noise, in this case, stabilizes the dynamics as can be seen from the reduction of average maximal Lyapunov exponent $\langle L_{max}\rangle$. This is due to the fact that the destroyed orbit has positive Lyapunov exponent.      

Another possible effect of noise when it is added to the system in a region of $\beta$ with coexistence of orbits, is to induce transitions between them (provided that both orbits are robust against noise). We have not observed this but a more comprehensive study should be make.

\begin{acknowledgments}
The authors wishes to acknowledge CONICET (PIP6124) for financial assistance. 
%\dots.
\end{acknowledgments}
 
\appendix*
\section{}
\label{sec:apendice}

%\section*{Appendix}
%\label{sec:apendice}
Taking in mind the future study of the quantum analogue of this kind of ratchet using the Bloch-Floquet formalism, we have considered the Fourier expansion of the Potential Eq.[\ref{eq:exacto}] rather than its exact expression. That is, for $N$ sites, $a=\frac{2\pi}{N}$ and $0 \leq x \leq 2\pi$:
\begin{equation}
\label{eq:fourier}
V_{\alpha}(x)=V_{o}\sum_{l=1} \left[ a_{l}(\alpha) cos(Nlx)+b_{l}(\alpha) sin(Nlx)\right].
\end{equation} 
The coeficients $a_{l}(\alpha)$ and $b_{l}(\alpha)$ have annalytical expressions given by:
\begin{eqnarray}
a_{l}(\alpha)= F_{l}(\alpha) \sin\left( \frac{2\alpha l \pi}{1+\alpha}\right),\nonumber \\
b_{l}(\alpha)=F_{l}(\alpha) \left[1+\cos\left( \frac{2\alpha l \pi}{1+\alpha}\right)\right];
\end{eqnarray}
where 
\begin{equation}
F_{l}(\alpha)=\frac {4 l (\alpha-1)(\alpha+1)^3}{[(\alpha+1)^2- (2l)^2][(2\alpha l)^2-(\alpha+1)^2]\pi } \; .
\end{equation}
For $(2l-1)=\alpha$ or $1/(2l-1)=\alpha$ these expessions are undeterminated but taking the appropiate limit, we obtain:
\begin{eqnarray}
a_{l}(\alpha)= \frac{1}{2 l},\nonumber \\
b_{l}(\alpha)=0.
\end{eqnarray}
As a mesure of the difference of expansion Eq.[\ref{eq:fourier}] up to 20 th order from the exact potential Eq.[\ref{eq:exacto}] we consider
\begin{equation}
\sigma=\frac{1}{V_{o}}\sqrt{\frac{1}{2\pi}\int_{0}^{2\pi} (V_{apr}-V_{exact})^{2} dx} = 6 \times 10^{-4}.
\end{equation}
In Figure \ref{fig:figuextra}(a), the exact potential is undistinguishable from its Fourier expansion.

\end {document}